\pdfoutput=1

\documentclass[reprint,aps,prl,showpacs,twocolumn]{revtex4-1}
\usepackage{mathptmx}
\usepackage{epsfig,amsopn}
\usepackage{graphicx}
\usepackage{amsmath,amssymb}
\usepackage{natbib}
\usepackage{xcolor}
\usepackage{braket}
\usepackage{float}
\usepackage{enumerate}
\usepackage{mathtools}
\usepackage[symbol]{footmisc}
\allowdisplaybreaks[1]

\newcommand{\eref}[1]{Eq.~(\ref{#1})}%

\def\bea{\begin{eqnarray}}
\def\eea{\end{eqnarray}}

\begin{document}

\title{P\'eclet number governs transition to acceleratory restart in drift-diffusion}

\author{{\normalsize{}Somrita Ray$^1$}{\normalsize{}}}

\author{{\normalsize{}Debasish Mondal$^2$}{\normalsize{}}}

\author{{\normalsize{}Shlomi Reuveni$^1$}{\normalsize{}}}
\email{Author for correspondence: shlomire@tauex.tau.ac.il}

\affiliation{\noindent \textit{
$^{1}$School of Chemistry, The Center for Physics and Chemistry of Living Systems, The Raymond and Beverly Sackler Center for Computational Molecular and Materials Science,\\ \& The Ratner Center for Single Molecule Science, Tel Aviv University, Tel Aviv 6997801, Israel\\
$^{2}$ Department of Chemistry, Indian Institute of Technology Tirupati, Tirupati 517506, India}}

\date{\today}

\begin{abstract}
\noindent
First-passage processes can be divided in two classes: those that are accelerated by the introduction of restart and those that display an opposite response. In physical systems, a transition between the two classes may occur as governing parameters are varied to cross a universal tipping point. However, a fully tractable model system to teach us how this transition unfolds is still lacking. To bridge this gap, we quantify the effect of stochastic restart on the first-passage time of a drift-diffusion process to an absorbing boundary. There, we find that the transition is governed by the P\'eclet number ($Pe$) --- the ratio between the rates of advective and diffusive transport. When $Pe>1$ the process is drift-controlled and restart can only hinder its completion. In contrast, when $0\leq~Pe<1$ the process is diffusion-controlled and restart can speed-up its completion by a factor of $\sim1/Pe$. Such speedup occurs when the process is restarted at an optimal rate  $r^{\star}\simeq r_0^{\star}\left(1-Pe\right)$, where $r_0^{\star}$ stands for the optimal restart rate in the pure-diffusion limit. The transition considered herein stands at the core of restart phenomena and is relevant to a large variety of processes that are driven to completion in the presence of noise. Each of these processes has unique characteristics, but our analysis reveals that the restart transition resembles other phase transitions --- some of its central features are completely generic. 
\end{abstract}

\maketitle
\section{I. Introduction}
Restart---a situation in which a dynamical process stops and starts anew---is a naturally occurring phenomenon as well as an integral part of many man-made systems. Random catastrophes can drastically reduce the population of a living species and restart its growth \cite{population1}. Computer algorithms are often restarted to shorten their total run times \cite{computerscience}, and financial crises may result in a crash of the stock market thus resetting asset prices to past levels \cite{economics}. Search processes may also undergo restart e.g., when bad weather forces teams to return to base \cite{PalReuveniPRL}, and when hunger or fatigue brings foraging animals back to their shelters \cite{forage}. At the microscopic level, restart is an integral part of the renowned Michaelis-Menten reaction scheme and is thus crucial to the understanding of enzymatic and heterogeneous catalysis \cite{Restart-Biophysics1, Restart-Biophysics2, Restart-Biophysics3, Restart-Biophysics4}. For these reasons and others, restart has recently attracted considerable attention.

Diffusion with stochastic resetting played a central role in shaping our understanding of restart phenomena \cite{D1,D2,D3,D4,D5,D6,D7,D8,D9,D10,D11}. In this model, one considers a Brownian particle that is returned to its initial position at random time epochs. This simple setup already gives rise to  non-equilibrium steady states, but the model can also be considered in the presence of an absorbing boundary to provide a quintessential example of a system where restart acts to facilitate the completion of a first-passage process \cite{FPT1,FPT2}. This counter-intuitive phenomenon is not unique to diffusion \cite{ReuveniPRL,PalReuveniPRL}. And yet, the fact that it can be demonstrated so vividly with a process that is known to physicists since the days of Einstein, Smoluchowski, and Perrin has really brought it to center stage and welcomed the advent of many important applications and generalizations  \cite{G&A1,G&A3,G&A4,G&A5,G&A6,G&A8,G&A9,G&A10,G&A11,G&A12,G&A13,G&A14,G&A15,G&A16,G&A17,G&A18,G&A19,G&A20,G&A21,G&A22,G&A23,G&A24,G&A25,G&A26}. In fact, some universal properties of first-passage under restart \cite{ReuveniPRL,PalReuveniPRL} were first demonstrated using diffusion. However, since restart always acts to expedite the first-passage time (FPT) of a free Brownian particle to a marked target, simple diffusion fails to display  a transition that is generically observed in many other systems.\\
\indent
Restart can either hinder or facilitate the completion of a stochastic process. Knowing the distribution of the random completion-time, one can tell---in advance---which of the two will happen \cite{ReuveniPRL,PalReuveniPRL,Restart-Biophysics1,Restart-Biophysics2}; but in real-world systems the answer will typically depend on environmental conditions. Indeed, changes in governing parameters such as temperature, viscosity, and concentration, can directly affect the progression of physical processes and chemical reactions, thus altering their stochastic completion-time distributions. When subject to such changes, restart may invert its role: hindering the completion of a process it previously facilitated and vice versa. This transition has already been predicted to have universal features \cite{ReuveniPRL,PalReuveniPRL}, as well as dramatic real-life implications \cite{Restart-Biophysics1,Restart-Biophysics2,Restart-Biophysics3,Restart-Biophysics4}, but an analytically tractable model system where it could be understood in full has so far been missing.\\
\begin{figure}[t]
\begin{centering}
\includegraphics[width=8.0cm]{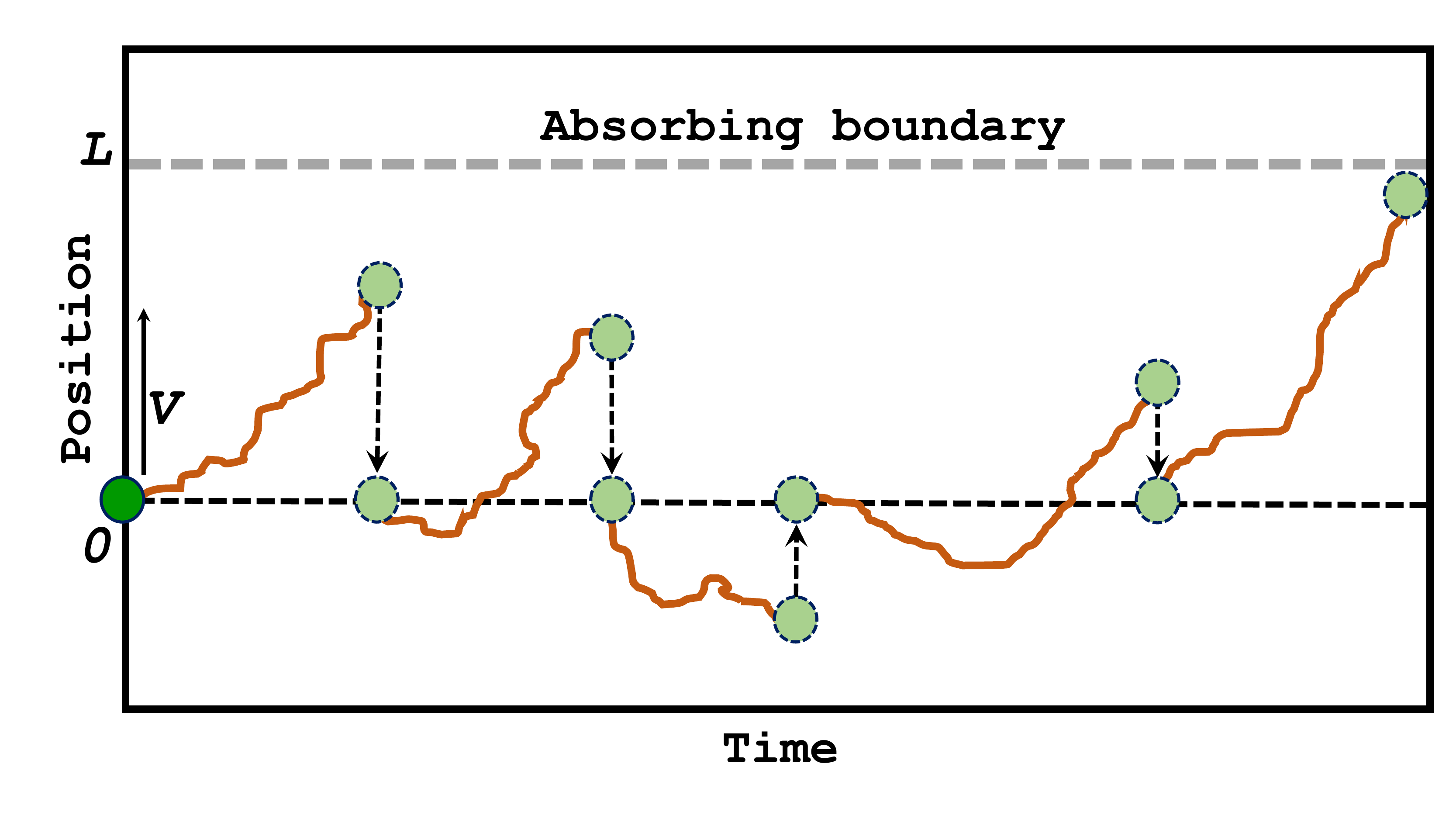}
\end{centering}
\caption{An illustration of drift-diffusion under stochastic restart.}
\label{Fig1}
\end{figure}
\indent
In many important situations diffusion occurs in the presence of a bias. This could happen, e.g., when a particle is diffusing in a flow-field or under the influence of some potential \cite{FP}. Sometimes, the particle under consideration is not even real but rather a symbolic entity that stands, e.g., for a reaction coordinate, the price of a stock, or the height of a fluctuating interface. This marked generality allows drift-diffusion to describe anything from charge transport \cite{CT} to biological evolution \cite{BE} and from polymer translocation \cite{PT1,PT2} to the dynamics of neuron firing \cite{NF}. Moreover, like many other first-passage processes \cite{PalReuveniPRL} drift-diffusion may also become subject to restart (Fig.~\ref{Fig1}); and it then provides a non-trivial example of a system where stochastic restart could either hinder or facilitate the first-passage of a particle to an absorbing boundary. A transition between the two types of behavior was shown to occur at a universal tipping point \cite{PalReuveni2}, but how this transition unfolds is still unknown. Here, we address this question and show that the answer to it reflects not only on the toy model illustrated in Fig.~\ref{Fig1}, but also on a large variety of processes that are driven to completion in the presence of noise. 

In this paper, we provide a comprehensive analysis of drift-diffusion under stochastic restart. We start in Sec. II where we utilize the Fokker-Planck approach to derive a detailed spatio-temporal description of the process illustrated in Fig.~\ref{Fig1}. A result for the FPT to the absorbing boundary follows and this is further corroborated using a general approach to first-passage under restart \cite{ReuveniPRL,PalReuveniPRL}. Both approaches are well-suited to treat generalizations of our problem, but aside from deductive reasons the latter is given to show that generalizations are not limited to processes described by the Fokker-Planck equation. In Sec. III, we demonstrate the restart transition in drift-diffusion and explain why it is reasonable to expect that it will be governed by the P\'eclet number ($Pe$) --- the ratio between the rates of advective and diffusive transport. In Sec. IV, we show that the P\'eclet number decides the effect of restart in our problem, and that it also sets the optimal restart rate, i.e., the rate which minimizes the mean FPT of the particle to the absorbing boundary. The speedup conferred by optimal restart is discussed in Sec. V, and it is shown that it too can be given in terms of $Pe$. In Sec. VI, we go on to show that the restart transition has a universal fingerprint on relative stochastic fluctuations of the FPT. In Sec. VII, we continue to discuss the generic nature of our results. The analogy with thermodynamic phase-transitions is then pointed out and discussed side-by-side with other broad implications of our findings and possible future extensions of our work. Final conclusions are given in Sec. VIII.  \\
\section{II. First-passage time of drift-diffusion under stochastic restart}
\subsection{A. Fokker-Planck approach}
Consider a particle undergoing drift-diffusion in one dimension and further assume that the particle has a constant diffusion coefficient $D$ and drift velocity $V$. This situation could e.g., arise for diffusion in a viscous medium under a linear potential $U(x)=-U_0x$, where $U_0$ is an arbitrary constant. Letting $\zeta$ stand for the drag coefficient, the drift velocity in this case will be $V=U_0/\zeta$, and the diffusion constant will be given by the Einstein-Smoluchowski relation $D=(\beta \zeta)^{-1}$ with $\beta$ standing for the thermodynamic beta.

To study the effect of resetting, we assume that the particle is further subjected to stochastic restart with a constant rate $r$, which means that it is returned to its initial position, $x=0$, after a random time that is taken from an exponential distribution with mean $1/r$. The process ends when the particle hits an absorbing boundary located at $x=L$. The master equation for $p(x,t)$, the conditional probability density of finding the particle at position $x$ at time $t$ provided its initial position was $x=0$, then reads
\begin{eqnarray}
\dfrac{\partial p(x,t)}{\partial t}=&&\nonumber \\
-V\dfrac{\partial p(x,t) }{\partial x}
&+&D\dfrac{\partial^2 p(x,t)}{\partial x^{2}}-r p(x,t)+r\delta(x)Q(t),
\label{eq:fme}
\end{eqnarray}
\noindent
where $\delta(x)$ is a Dirac delta function and $Q(t)\coloneqq\int_{-\infty}^{L}p(x,t)dx$ denotes the survival probability, i.e., the probability that the particle is not absorbed at the boundary by time $t$. We note that when the particle hits the boundary it is immediately absorbed, which sets the boundary condition to $p(L,t)=0$. In the absence of restart ($r=0$), Eq.~(\ref{eq:fme}) boils down to the Fokker-Planck description of a simple drift-diffusion process. When subjected to stochastic restart, the process experiences a loss of probability from position $x$ and a gain of probability at $x=0$, where the particle restarts its motion. The last two terms on the right hand side of Eq.~(\ref{eq:fme}) account for this additional probability flow. \\
\indent
Calculating the survival probability, $Q(t)$, is our principal objective here as this function holds all the information on the distribution of the FPT of the particle to the boundary, but in what follows we will also solve for $p(x,t)$. To do this, we take the Laplace transform of Eq.~(\ref{eq:fme}) and rearrange to obtain
\begin{eqnarray}
\dfrac{\partial^2 \tilde{p}(x,s)}{\partial x^2}-\left(\frac{V}{D}\right)\dfrac{\partial \tilde{p}(x,s)}{\partial x}&-&\left(\frac{s+r}{D}\right)\tilde{p}(x,s) \nonumber \\
&=&\frac{-1-r \tilde{Q}(s)}{D}\delta(x),
\label{eq:lt}
\end{eqnarray}
where $\tilde{p}(x,s)\coloneqq \int_0^{\infty}e^{-st}p(x,t)dt$ and $\tilde{Q}(s)\coloneqq \int_0^{\infty}e^{-st}Q(t)dt$ denote the Laplace transforms of $p(x,t)$ and $Q(t)$, respectively.
\noindent
The above differential equation can be solved exactly [see Appendix A] to give 
 \begin{equation}
\tilde{p}(x,s)=
\begin{cases}
\frac{1+r \tilde{Q}(s)}{D (\alpha_--\alpha_+)}\left[e^{\alpha_+ (x-L)+\alpha_-\;L}-e^{\alpha_- x}\right]&\text{if}\;\;\;\;0\leq x\leq\;L\\ \\
\frac{1+r \tilde{Q}(s)}{D (\alpha_--\alpha_+)}\left[e^{\alpha_+ (x-L)+\alpha_-\;L}-e^{\alpha_+ x}\right]&\text{if}\;-\infty<x\leq\;0,\\
\end{cases}
\label{eq:pxs}
\end{equation}
where $\alpha_{\pm} =\frac{1}{2D}\left[V\pm\sqrt{V^2+4D(s+r)}\right]$.\\
\indent
The result in Eq.~(\ref{eq:pxs}) gives $p(x,t)$ in terms of $Q(t)$ in Laplace space. To get the Laplace transform of $Q(t)$ in a self-consistent manner from the above, we observe that $\tilde{Q}(s)\coloneqq\int_{-\infty}^{L}\tilde{p}(x,s)dx$. Integrating Eq.~(\ref{eq:pxs}) over the spatial coordinate and rearranging we find
\begin{equation}
\tilde{Q}(s)=\frac{1-\exp{\left[{\frac{L}{2D}\left(V-\sqrt{V^2+4D(s+r)}\right)}\right]}}{s+r\; \exp{\left[{\frac{L}{2D}\left(V-\sqrt{V^2+4D(s+r)}\right)}\right]}}.\\
\label{eq:qls_sol}
\end{equation}
\noindent
Note that the conventional approach to first-passage usually starts with the backward Fokker-Planck description of the system \cite{gard}. This leads directly to the survival probability, bypassing the calculation of the spatio-temporal probability distribution function. In contrast, by taking the forward Fokker-Planck approach, we can plug Eq.~(\ref{eq:qls_sol}) back into Eq.~(\ref{eq:pxs}) and obtain an explicit expression for the propagator, $\tilde{p}(x,s)$, in Laplace space. The forward Fokker-Planck approach thus helped us extract additional information regarding the position of the particle and this may be useful in future studies.\\
\indent
Eq.~(\ref{eq:qls_sol}) allows us to compute the FPT of the particle to the absorbing boundary. Letting $T_r$ denote this FPT, we recall that the probability density function of this random variable is given by $-d Q(t)/dt$ \cite{gard}. This allows us to calculate any moment of $T_r$ from $\tilde{Q}(s)$ following the relation
\begin{equation}
\left<T_r^n\right>=\left(-1\right)^{n-1}n\left[\frac{d^{n-1}\tilde{Q}(s)}{ds^{n-1}}\right]_{s=0}.\\
\label{eq:fpt_moments}
\end{equation}
The above relation will be useful in calculating the standard deviation, $\sigma(T_r)=\sqrt{\left<T_r^2\right>-\left<T_r\right>^2}$ in Sec. VI. Here we focus on the the first moment, i.e., the mean FPT, which is given by $\left<T_r\right>=\left[\tilde{Q}(s)\right]_{s=0}$. Thus from Eq.~(\ref{eq:qls_sol}) we obtain
\begin{equation}
\left<T_r\right>=\frac{\exp{\left[{\frac{L}{2D}\left(\sqrt{V^2+4Dr}-V\right)}\right]}-1}{r}.\\
\label{eq:mfpt}
\end{equation}
Eq.~(\ref{eq:mfpt}) relates the mean FPT of a drift-diffusion process under restart with the fundamental physical parameters that govern this problem, viz. $V$, $D$, $L$ and $r$. In the next subsection, we derive the mean and the distribution of $T_r$ following a general approach to first-passage under restart. 
\subsection{B. General approach}
\indent
A general theory of first-passage under restart was developed in \cite{ReuveniPRL,PalReuveniPRL}. This theory asserts that one can write the FPT distribution of a process that is restarted at a rate $r$ in terms of the FPT distribution of the process without restart. More specifically, letting $T$ denote the FPT of some generic process and $T_r$ its FPT under restart, we define $\tilde{T}(s)\coloneqq\left<\exp{(-sT)}\right>$ and $\tilde{T}_r(s)\coloneqq\left<\exp{(-sT_r)}\right>$ to be the Laplace transforms of these random variables. The following relation then holds \cite{ReuveniPRL} 
\begin{equation}
\tilde{T}_r(s)=\frac{\tilde{T}(s+r)}{\frac{s}{s+r}+\frac{r}{s+r} \tilde{T}(s+r)}.\\
\label{eq:fptd_r}
\end{equation}
The mean FPT of a generic process under stochastic restart follows directly from \eref{eq:fptd_r} and is given by \cite{ReuveniPRL} 
\begin{equation}
\left<T_r\right>=\frac{1}{r}\left[\frac{1-\tilde{T}(r)}{\tilde{T}(r)}\right].\\
\label{eq:mfpt_lt}
\end{equation}
Eqs.~(\ref{eq:fptd_r}) and (\ref{eq:mfpt_lt}) are completely general. To demonstrate this, we will now use them to analyze drift-diffusion under stochastic restart.

The first-passage time $T$ of a drift-diffusion process to an absorbing boundary is known to be governed by the following probability density function \cite{Cox-Miller-Book}
\begin{equation}
f_T(t)=\frac{L}{\sqrt{4\pi Dt^3}} \exp{\left[-\frac{(L-Vt)^2}{4Dt}\right]},\\
\label{eq:fpt_dist}
\end{equation}
\noindent
where once again we have $L$, $D$, and $V$ standing respectively for the initial distance from the boundary, the diffusion coefficient, and the drift velocity. The Laplace transform of $T$ is also known exactly and is given by  
\begin{equation}
\tilde{T}(s)=\exp{\left[\frac{L}{2D}\left(V-\sqrt{V^2+4Ds}\right)\right]}. \\
\label{eq:fpt_lt}
\end{equation}
Plugging in Eq.~(\ref{eq:fpt_lt}) into Eq.~(\ref{eq:mfpt_lt}), one readily recovers  Eq.~(\ref{eq:mfpt}) for the mean FPT, $\left<T_r\right>$. Plugging in Eq.~(\ref{eq:fpt_lt}) into Eq.~(\ref{eq:fptd_r}), we find
\begin{equation}
\tilde{T}_r(s)=\frac{s+r}{r+s\;\exp\left({\frac{L}{2D}\left[\sqrt{V^2+4D(s+r)}-V\right]}\right)}.\\
\label{eq:fptd_r_dd}
\end{equation}
To show that this result is equivalent to the one in \eref{eq:qls_sol} we recall that $Q(t)=\int_{t}^{\infty}f_{T_r}(t)dt$, where $f_{T_r}(t)$ is the FPT distribution of drift-diffusion under restart; hence $\tilde{Q}(s)=[1-\tilde{T_r}(s)]/s$. Equivalence of Eqs. (\ref{eq:qls_sol}) and (\ref{eq:fptd_r_dd}) is then evident. 

\section{III. Restart transition}
\begin{figure}[t!]
\begin{centering}
\includegraphics[width=8.0cm]{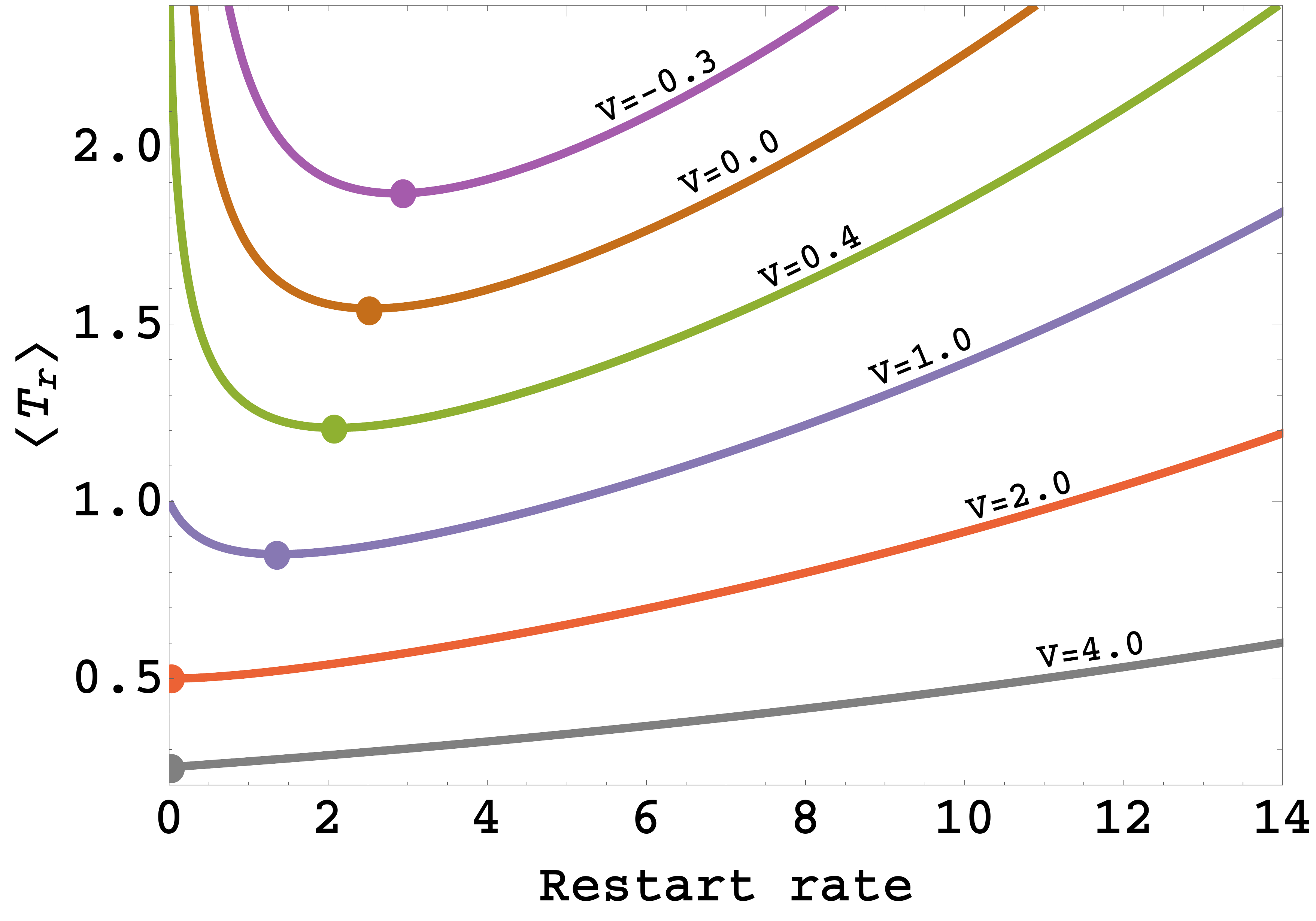}
\end{centering}
\caption{The mean FPT from Eq.~(\ref{eq:mfpt}) vs. the restart rate for different values of the drift velocity $V$. Here, $D=1$ and $L=1$. Circles indicate minima.}
\label{Fig2}
\end{figure}

\indent
Having obtained an explicit expression for the mean FPT of a drift-diffusion process under stochastic restart [Eq.~(\ref{eq:mfpt})], we now turn to explore how this depends on physical parameters. In Fig.~\ref{Fig2}, we plot the mean FPT, $\left<T_{r}\right>$, vs. the restart rate, $r$ for different values of the drift velocity $V$. For $V=0$, i.e., pure diffusion, we see that the mean FPT behaves non-monotonically with $r$, obtaining a minimum at an optimal restart rate $r=r^{\star}$, as was first observed by Evans and Majumdar in \cite{D1}. A similar behavior is also observed when the drift velocity is positive, but sufficiently small. However, at some point, as $V$ increases, a transition occurs, and $\left<T_{r}\right>$ becomes monotonically increasing with $r$. This transition can also be identified by examining the optimal restart rate. This is maximal for pure diffusion, and gradually decreases as $V$ increases until it vanishes at some critical drift velocity. When $V<0$, $\left<T_{r}\right>$ is non-monotonic [Fig. \ref{Fig2}] and the optimal restart rate increases with the magnitude of the (negative) drift velocity. In what follows, we focus on the $V>0$ regime where a transition in the role of restart occurs. The $V<0$ regime is analyzed separately in Appendix B. \\ 
\indent
The transition seen in Fig.~\ref{Fig2} can be understood from Eq.~(\ref{eq:mfpt}) by contrasting the pure-diffusion and pure-drift ($V>0$) limits. For pure diffusion \cite{D1}, substituting $V=0$ in Eq.~(\ref{eq:mfpt}) we get $\left<T_{r}\right>=\left[\exp{\left(\sqrt{rL^2/D}\right)}-1\right]/r$. Then it is easy to see that $\left<T_{r}\right>$ diverges for the very high and very low values of $r$, but is finite otherwise. Thus one expects the mean FPT to have a minimum with respect to the restart rate. In the other limit, when $V\rightarrow\infty$, one can approximate the mean FPT as $\left<T_{r}\right>\simeq\left[\exp{\left(rL/V\right)}-1\right]/r$. Taylor expanding the exponential term, we find that the mean FPT attains the value $L/V$ at $r=0$ and increases monotonically with $r$ from that point onward. Recapitulating, we see that for high values of $r$, $\left<T_{r}\right>$ diverges in all cases, because frequent resetting makes it harder for the particle to hit the absorbing boundary. In contrast, the behavior at low values of $r$ is sensitive to the drift velocity. This analysis suggests that the mean FPT should generally show a non-monotonic variation at the diffusion-controlled regime and a monotonic increase at the drift-controlled regime.  \\
\indent
A standard way to compare drift and diffusion is by use of the P\'eclet number which is defined as the ratio between the rates of advective and diffusive transport  
\begin{equation}
Pe\coloneqq~LV/2D.\\
\label{eq:Pe}
\end{equation}
The P\'eclet number naturally appears in classic first-passage time problems that involve drift-diffusion \cite{FPT1}. In the absence of drift, the above definition makes it clear that $Pe=0$. As the drift velocity increases, or as the diffusion coefficient decreases, the P\'eclet number increases. In other words, low values of $Pe$ correspond to a diffusion-controlled regime, whereas high values of $Pe$ correspond to a drift-controlled regime. In this way, the P\'eclet number beautifully captures the interplay between drift and diffusion, and it thus makes sense to try and characterize the transition seen in Fig.~\ref{Fig2} in terms of this dimensionless quantity. Before moving on, we note that for $V<0$ the P\'eclet number is negative by definition and refer to Appendix B for analysis of this case. 
\indent
\section{IV. Optimal restart}
The theory of first-passage under restart asserts that the introduction of stochastic restart will result in a decrease of the mean FPT whenever the ratio between the standard deviation and mean of the FPT distribution---in the absence of restart---is larger than unity, and vice versa \cite{Restart-Biophysics1,Restart-Biophysics2,ReuveniPRL,PalReuveniPRL}. In our case, the mean and standard deviation of the underlying first-passage process can be calculated directly from Eq.~(\ref{eq:fpt_dist}) to give: $\left<T\right>=L/V$ and $\sigma(T)=\sqrt{2DL/V^3}$. Their ratio, the coefficient of variation ($CV$), is then given by $CV=1/\sqrt{Pe}$. It is thus clear that $Pe=1$ marks the point of transition, but the general theory tells us nothing about how this transition unfolds. To answer this question, we will now examine the optimal restart rate $r^{\star}$ and show that it too can be expressed in terms of $Pe$.\\

\indent
We start by defining a reduced variable   
\begin{equation}
z\coloneqq{\frac{L}{2D}\left[\sqrt{V^2+4Dr}-
V\right]},\\
\label{eq:z_def}
\end{equation}
and rewriting the restart rate $r$ in terms of $z$ to give
\begin{equation}
r=\left(\frac{V}{L}\right)z+\left(\frac{D}{L^2}\right)z^2.\\
\label{eq:r_z}
\end{equation}
\noindent
Substituting Eq.~(\ref{eq:r_z}) into Eq.~(\ref{eq:mfpt}), we express the mean FPT in terms of $z$ as
\begin{equation}
\left<T_r\right>=\left(\frac{L^2}{LV+Dz}\right)\left[\frac{\exp{(z)}-1}{z}\right].\\
\label{eq:Tr_z}
\end{equation}

\begin{figure}[t!]
\begin{centering}
\includegraphics[width=8.0cm]{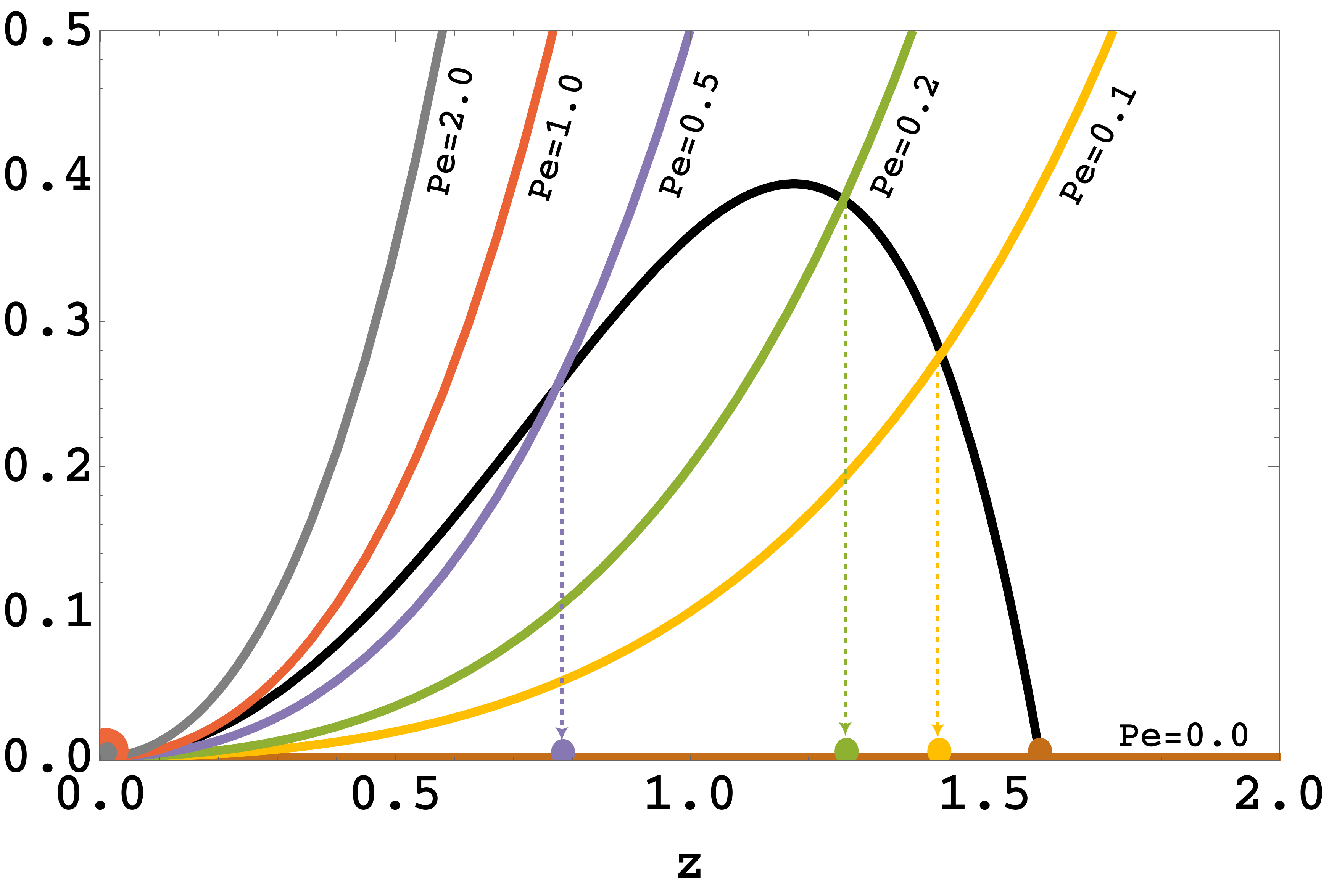}
\end{centering}
\caption{Plots of the left-hand side (colored lines) and right-hand side (black line) of Eq.~(\ref{eq:trans}) vs. the reduced variable $z$ from \eref{eq:z_def}. Colored circles denote solutions to \eref{eq:trans} for different P\'eclet numbers.}
\label{Fig3}
\end{figure}

To find the optimal restart rate, we now look for a solution to
\begin{equation}
\frac{d\left<T_r\right>}{dr}=\frac{1}{LV+2Dz}\left[\frac{d\left<T_r\right>}{dz}\right]=0.\\
\label{eq:Tr_min}
\end{equation}
Substituting Eq.~(\ref{eq:Tr_z}) into Eq.~(\ref{eq:Tr_min}), we obtain the following transcendental equation 
\begin{equation}
\left[\left(z-1\right)\exp{(z)}+1\right]Pe=\left[\left(1-\frac{z}{2}\right)\exp{(z)}-1\right]z.\\ 
\label{eq:trans}
\end{equation}
\noindent
We now solve Eq.~(\ref{eq:trans}) and calculate the optimal restart rate in terms of its solutions. In Fig.~{\ref{Fig3}}, we simultaneously plot the left hand side, $F_1(z,Pe):=\left[\left(z-1\right)\exp{(z)}+1\right]Pe$ (colored lines), and the right hand side, $F_2(z):=\left[\left(1-\frac{z}{2}\right)\exp{(z)}-1\right]z$ (black line), of Eq.~(\ref{eq:trans}). The solutions, denoted as $z^{\star}$, are the $z$ values for which $F_1(z,Pe)$ and $F_2(z)$ intersect. We observe that for $Pe<1$, Eq.~(\ref{eq:trans}) has one non-trivial positive solution $z^{\star}>0$. In the pure-diffusion limit, $Pe=0$, and $z^{\star}$ attains its maximal value $z_0^{\star}\simeq1.59$. It then gradually decreases as $Pe$ increases until it vanishes for $Pe\geq1$.
\indent
Following Eq.~(\ref{eq:r_z}), and the solution of Eq.~(\ref{eq:trans}), the optimal restart rate for a drift-diffusion process can then be written as
\begin{equation}
\frac{r^{\star}}{r_0^{\star}}=\frac{2~Pe~z^{\star}+z^{{\star}^2}}{z_0^{{\star}^2}},\\
\label{eq:optr_Pe}
\end{equation}
where $r_0^{\star}=Dz_0^{{\star}^2}/L^2$ stands for the optimal restart rate in the pure diffusion limit \cite{D1}. Recalling that $z^{\star}$ is uniquely determined by the P\'eclet number, we see from Eq.~(\ref{eq:optr_Pe}) that the same can be said about the scaled optimal restart rate: $r^{\star}/r_0^{\star}$. 
\\

\begin{figure}[t]
\begin{centering}
\includegraphics[width=8.4cm]{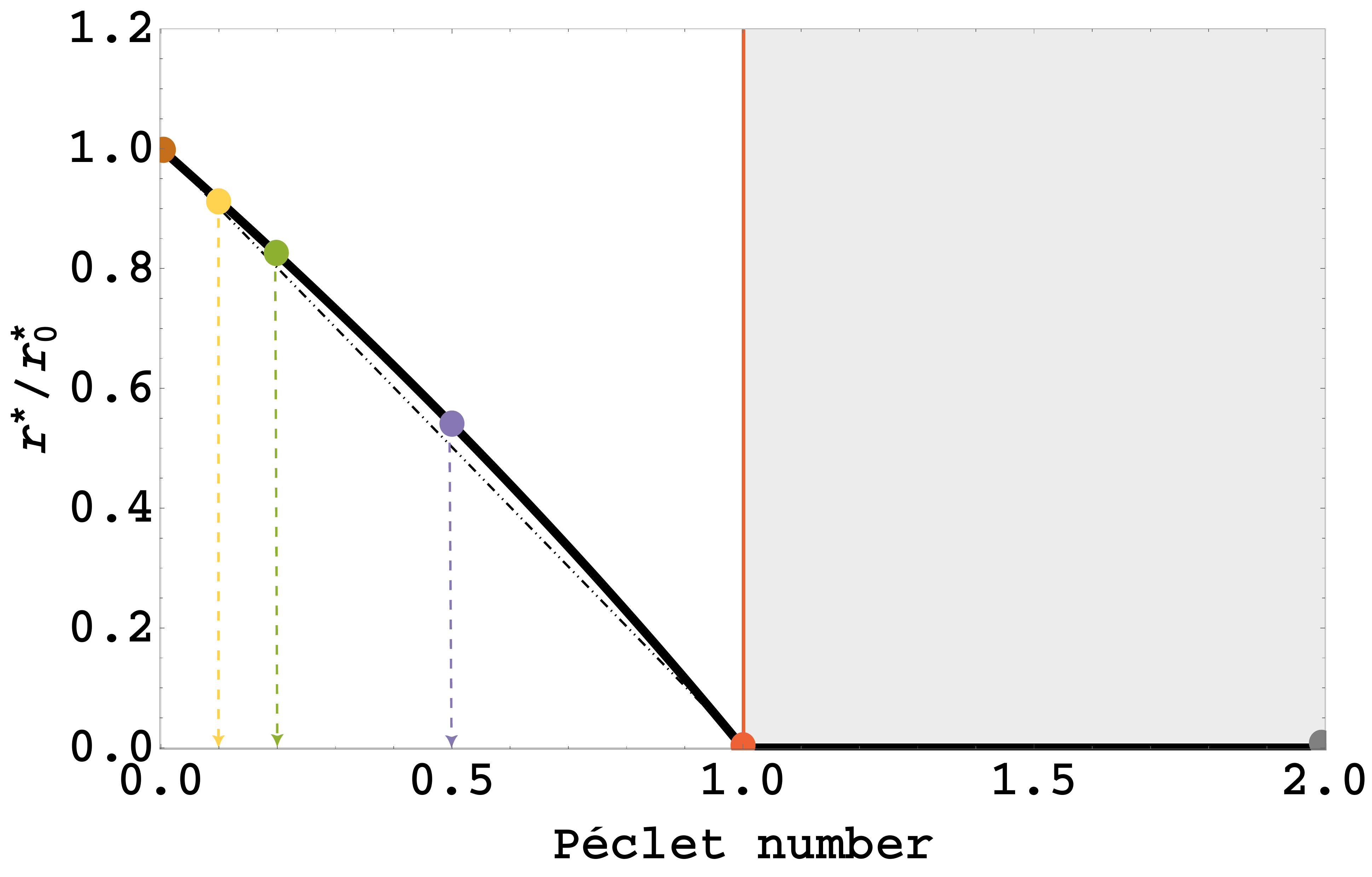}
\end{centering}
\caption{The scaled optimal restart rate from \eref{eq:optr_Pe} vs. the P\'eclet number from \eref{eq:Pe} (solid black line). A transition  occurs at $Pe=1$ (solid red line). Colored circles correspond to $Pe$ values from Fig.~\ref{Fig3}. The linear approximation from  Eq.~(\ref{eq:scale_r}) is marked as a dot-dashed black line.}
\label{Fig4}
\end{figure}

\indent
In Fig.~\ref{Fig4}, we present the scaled optimal restart rate vs. the P\'eclet number. We observe that optimal restart rates are strictly positive for $Pe<1$ (white region), which means that restart speeds up the first-passage process in the diffusion-controlled regime. On the contrary, optimal restart rates are always zero for $Pe\geq1$ (gray region), which means that restart does not speed up  first-passage in the drift-controlled regime. This clearly indicates a transition at $Pe=1$. Finally, we graphically observe that for $Pe\leq1$ the scaled optimal restart rates exhibit an almost linear dependence on the P\'eclet number  
\begin{equation}
r^{\star}/r_0^{\star}\simeq\left(1-Pe\right).
\label{eq:scale_r}
\end{equation}
\noindent
\eref{eq:scale_r} provides a simple and effective way to approximate the optimal restart rate for $Pe\leq1$. In Sec. VII, we utilize this expression in order to explore how temperature affects the restart transition.

\section{V. Maximal speedup}
\indent
The mean FPT of a drift-diffusion process with a constant drift velocity $V$, diffusion coefficient $D$, and initial distance $L$ from an absorbing boundary, can be readily calculated from the FPT distribution in \eref{eq:fpt_dist} to give $\left<T\right>=L/V$, and note that this result does not depend on the diffusion coefficient. When the same drift-diffusion process is subjected to optimal restart, its mean FPT can be expressed as
\begin{equation}
\left<T_{r^{\star}}\right>=
\begin{cases}
\frac{L^2}{\left(LV+Dz^{{\star}}\right)}\left[\frac{\exp{(z^{\star})}-1}{z^{\star}}\right] & \text{if}\;\;\;0\leq\;Pe<1\\
\frac{L}{V} & \text{if}\;\;\;Pe\geq\;1,\\
\end{cases}
\label{eq:mfpt_opt}
\end{equation}
where we substituted $z$ in \eref{eq:Tr_z} by $z^{\star}$ and recalled that $z^{\star}=0$ for $Pe\geq1$. The speedup optimal restart confers on the mean FPT of a drift-diffusion process then reads
\begin{equation}
\frac{\left<T\right>}{\left<T_{r^{\star}}\right>}=
\begin{cases}
\frac{z^{{\star}^2}+2\;Pe\;z^{\star}}{2\;Pe\;\left(\exp{(z^{\star})}-1\right)} & \text{if}\;\;\;0\leq\;Pe<1\\
1 & \text{if}\;\;\;Pe\geq\;1.\\
\end{cases}
\label{eq:speedup}
\end{equation}
\noindent
Equation (\ref{eq:speedup}) shows that the maximal speedup, just like the scaled optimal restart rate, can be uniquely characterized by the P\'eclet number. 

When $Pe\ll1$, $z^{\star}\simeq\;z_0^{\star}$, suggesting a simple scaling law for the maximal speedup
\begin{equation}
\left<T\right>/\left<T_{r^{\star}}\right>\sim\;1/Pe.
\label{eq:speedup_scale} 
\end{equation}
To understand the importance of Eq.~(\ref{eq:speedup_scale}), consider drift-diffusion without restart. There, the mean FPT to the boundary is given by $\left<T\right>=L/V$ which asserts that doubling the drift velocity will half the mean time to completion. In contrast, Eq.~(\ref{eq:speedup_scale}) asserts that, when $Pe$ is sufficiently small, optimal restart will expedite completion many folds more. This point is further discussed in Sec. VII. 

In Fig.~\ref{Fig5}, we plot $\left<T\right>/\left<T_{r^{\star}}\right>$ as a function of the P\'eclet number to show that the maximal speedup diverges for a pure diffusion process ($Pe=0$), and that it monotonically decays to unity as $Pe\rightarrow1$. This figure also shows that when $Pe\ll1$ the speedup is inversely proportional to the P\'eclet number. Finally, it is evident from Fig.~\ref{Fig5} that for $Pe\geq1$ there is no speedup, which is another way of saying that the optimal restart rate is zero in the drift-controlled regime.  
\section{VI. Fluctuations}
\indent
Having fully characterized the transition on the level of the mean FPT, we now turn to look at stochastic fluctuations. Eq.~(\ref{eq:fpt_moments}) allows us to calculate all the moments of the FPT. In particular, we find that the standard deviation of the FPT is given by
\begin{figure}[t]
\begin{centering}
\includegraphics[width=8.4cm]{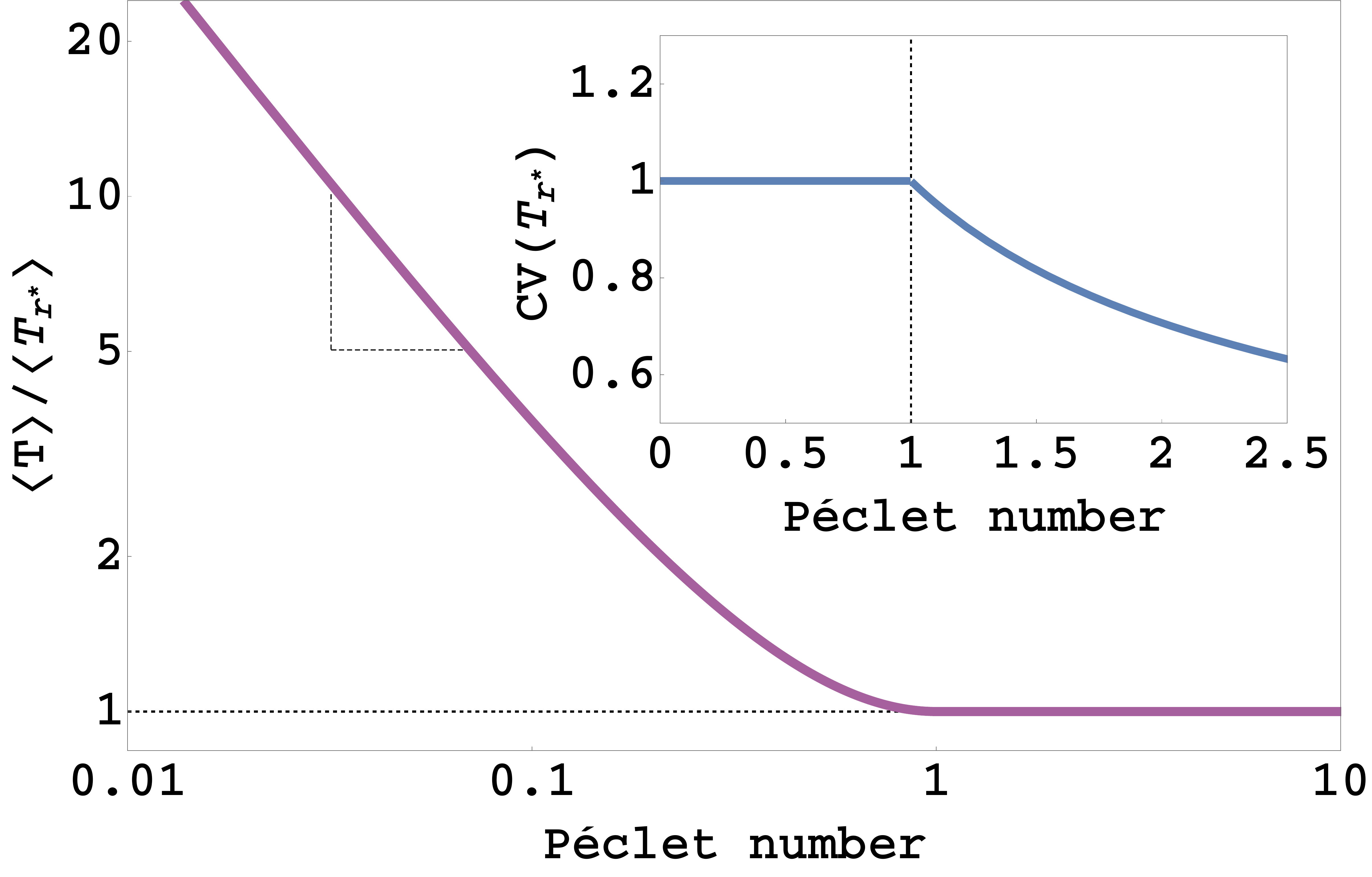}
\end{centering}
\caption{Main: The maximal speedup conferred by optimal restart [\eref{eq:speedup}] vs. the P\'eclet number from \eref{eq:Pe}. The scaling law from \eref{eq:speedup_scale} is clearly visible. Inset: The coefficient of variation under optimal restart from \eref{eq:CV_opt} vs. the P\'eclet number. In both the inset and main panel a clear transition is seen at $Pe=1$.}
\label{Fig5}
\end{figure}
\begin{align}
\sigma(T_r)=\frac{\sqrt{\exp{(2z)}-2\exp{(z)}\frac{Lr}{\sqrt{4
Dr+V^2}}-1}}{r},
\label{eq:stdv}
\end{align}
where $z$ is defined in  Eq.~(\ref{eq:z_def}). The coefficient of variation, $CV(T_r)=\sigma(T_r)/\left<T_r\right>$, is a dimensionless measure of stochastic fluctuations in the FPT. Combining Eqs.~(\ref{eq:mfpt}) and (\ref{eq:stdv}), we obtain
\begin{equation}
CV(T_r)=\frac{\sqrt{\exp{(2z)}-2z\exp{(z)}\frac{Pe+z/2}{Pe+z}-1}}{\exp{(z)}-1}.\\
\label{eq:cv}
\end{equation}
Eq.~(\ref{eq:cv}) holds for any restart rate $r$. To demonstrate a transition, we once again focus on the optimal restart rate $r^{\star}$. Recalling that $r^{\star}>0$ is a solution to Eq.~({\ref{eq:trans}}) for $Pe<1$, we rearrange this equation to get: $z\exp{(z)}(Pe+z/2)/(Pe+z)=\exp{(z)}-1$, and substitute back into \eref{eq:cv} to obtain  $CV(T_{r^{\star}})=1$. For $Pe\geq~1$, however, $r^{\star}=0$, i.e., the process is nothing but the underlying drift-diffusion process itself. In this regime, $CV(T_{r^{\star}})=CV(T)=1/\sqrt{Pe}$ as we discussed earlier in Sec. IV. Summing up, we have
\begin{equation}
CV(T_{r^{\star}})=
\begin{cases}
1 & \text{if}\;\;\;0\leq\;Pe<1\\
\frac{1}{\sqrt{Pe}} & \text{if}\;\;\;Pe\geq\;1.\\
\end{cases}
\label{eq:CV_opt}
\end{equation}
The transition in \eref{eq:CV_opt} is graphically illustrated in the inset of Fig.~\ref{Fig5}. Importantly, it should be noted that this transition is a concrete manifestation of a more general result: $CV(T_{r^{\star}})=1$ for any FPT process that is restarted at an optimal rate $r^{\star}>0$ \cite{ReuveniPRL}. On the other hand, when the optimal restart rate is zero, $CV(T_{r^{\star}})=CV(T)$. This quantity then depends on the details of the underlying first-passage process and is clearly non-universal.  
\section{VII. Discussion}
\begin{figure}[t]
\begin{centering}
\includegraphics[width=8.0cm]{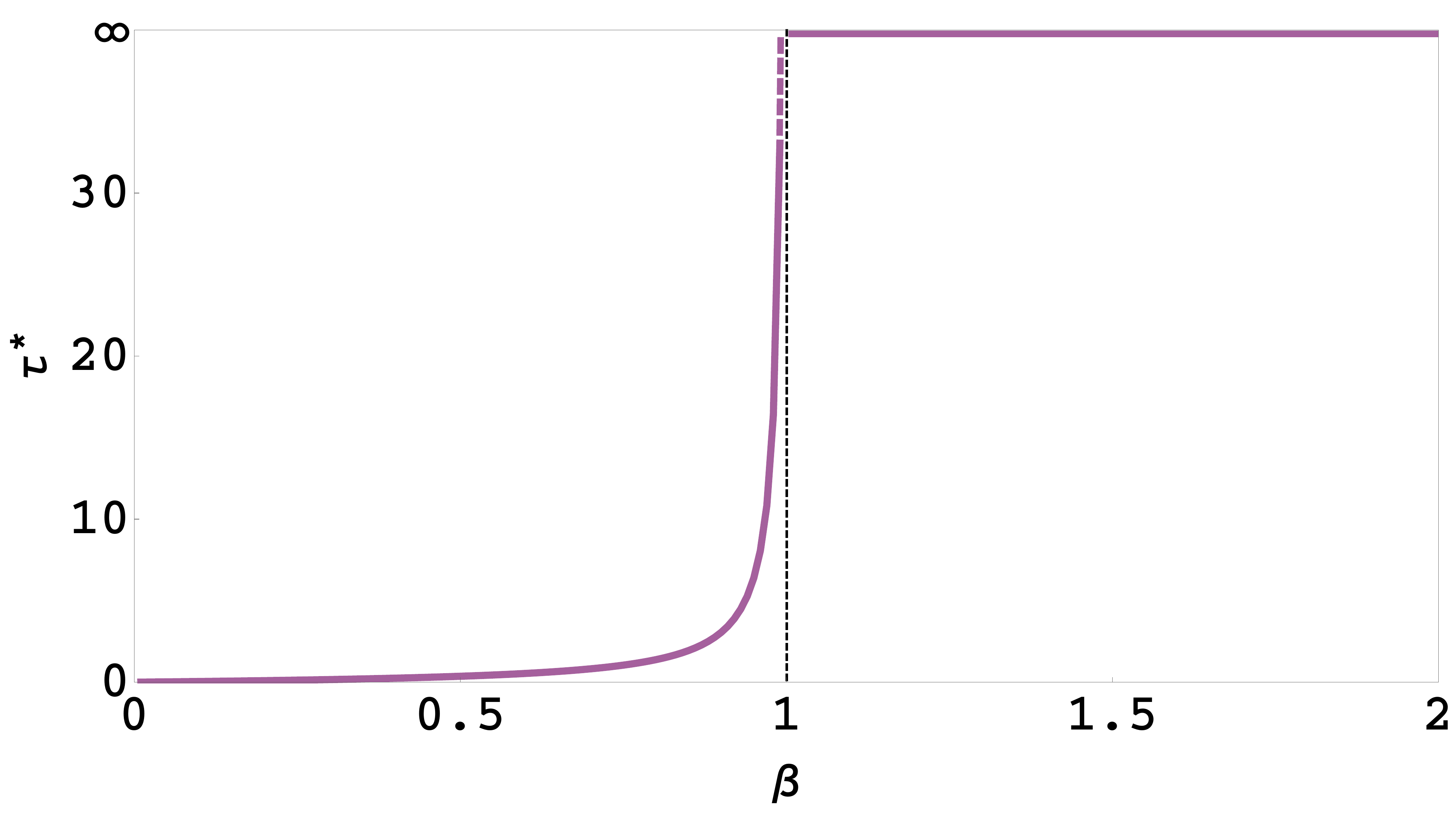}
\end{centering}
\caption{The mean optimal time between restart events, $\tau^{\star}=1/r^{\star}$, diverges at a critical temperature  $\beta_c^{-1}$ as described by  \eref{eq:opt_time_scal}. This generic feature of the restart transition is akin to that observed in other phase transitions. Plot was made using \eref{eq:optr_Pe} with parameters chosen such that $\beta_c=1$.} 
\label{Fig6}
\end{figure}

\begin{figure}[t!]
\begin{centering}
\includegraphics[width=8.2cm]{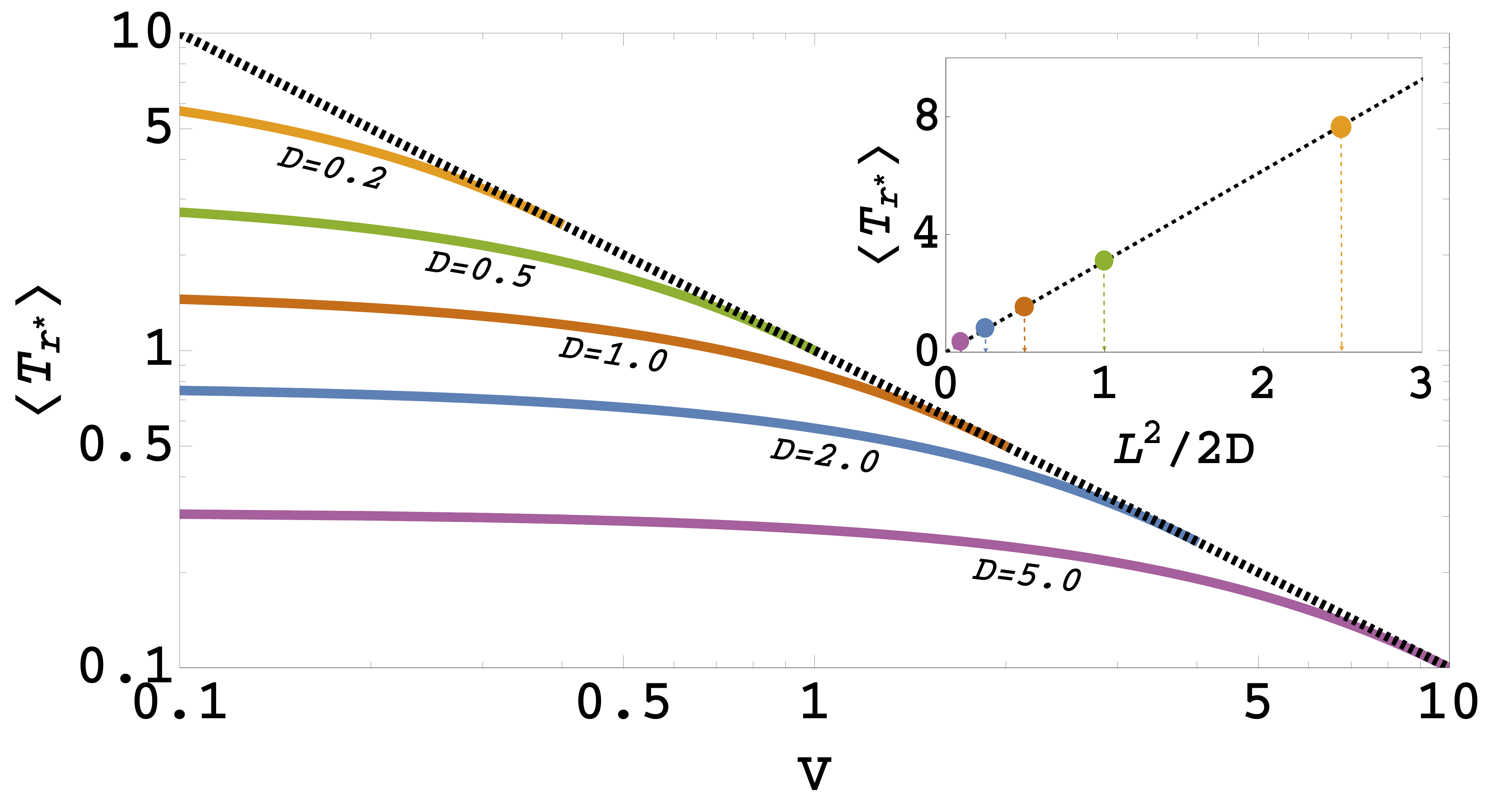}
\end{centering}
\caption{Main: The optimal mean FPT from Eq.~({\ref{eq:mfpt_opt}}) vs. the drift velocity $V$ for different values of the diffusion coefficient $D$ (colored lines). The mean FPT in the absence of any restart is also indicated (dashed black line). In all plots, $L=1$.
Inset: The optimal mean FPT for pure diffusion vs. $L^2/2D$. Circles correspond to asymptotic, $V \to 0$, values of the mean FPT in the main panel.}
\label{Fig7}
\end{figure}

A constant drift velocity emerges when a particle diffuses under a linear potential, e.g., the potential considered herein, $U(x)=-U_0x$; and note that when $U_0>0$ the particle achieves its minimum potential energy at the absorbing boundary ($x=L$). More complicated potentials can also be considered, but note that if the potential $U(x)$ has a single minimum at the absorbing boundary a restart transition is expected in the generic case. This transition will occur, e.g., as the diffusion constant is ramped up from an initial value of zero or ramped down from a very high value. Recalling that $D=(\beta \zeta)^{-1}$, we note that this could be achieved by heating or cooling the system, thus suggesting a critical transition temperature $\beta_c^{-1}$. For the model analyzed in this paper, we find that $\beta_c^{-1}=LV\zeta/2$. Moreover, starting from \eref{eq:scale_r}, we find that as the system is cooled down from a high temperature the transition gives rise to a diverging time scale 
\begin{equation}
\tau^{\star} = 1/r^{\star} \propto \bigg( \frac{\beta_c}{\beta} -1 \bigg)^{-1},\\
\label{eq:opt_time_scal}
\end{equation}
which governs optimal restart in the limit  $\beta\rightarrow\beta_c$ (Fig. \ref{Fig6}). The analogy with thermodynamic phase transitions is striking and raises the question whether the critical exponent in \eref{eq:opt_time_scal} is universal. Some evidence in support of this conjecture have recently been given in \cite{exponent}, but a general and formal proof still awaits to be found.\\
\indent
Our analysis reveals another generic feature of restarted processes that are driven to completion in the presence of noise. To illustrate this, we plot the mean FPT, $\left<T\right>=L/V$, of a drift diffusion process in the absence of restart vs. the drift velocity (Fig. \ref{Fig7}, dashed black line). The mean FPT under optimal restart is also plotted for different values of the diffusion coefficient (Fig. \ref{Fig7}, colored lines). The drift velocity serves here as a proxy for the drive strength, and one can trivially observe that increasing the drive always results in a lower mean FPT. Moreover, if the process is driven hard---so as to overwhelm noise---restart will not carry any benefit and should be avoided. However, in reality, one's ability to drive a process towards a desired outcome is typically limited due to practical constraints, and restart may then come to the rescue.  \\
\indent
Fig. \ref{Fig7} shows that restart can effectively surrogate for drive in the high-noise/low-drive limit. Moreover, in this limit optimal restart renders the mean FPT insensitive to the drive strength. Thus, when the application of drive is costly, complicated, or hard (in comparison to restart) it is best not to drive the system at all and restart it instead. This is clearly visible from the plots which show results for simple drift-diffusion, but it is important for us to emphasize that the behaviour seen here is totally generic. Namely, it does not depend on the details of the potential $U(x)$ that drives the particle. Indeed, when noise is very strong one can forget about the potential altogether and treat the process as if it were purely diffusive. Optimal mean FPTs then follow the familiar $\sim L^2/2D$ scaling (Fig. \ref{Fig7}, inset) which governs pure diffusion with resetting \cite{D1}.    
\section{VIII. Conclusions and Outlook}
In this paper, we presented a thorough analysis of drift-diffusion under stochastic restart. We calculated the first-passage time of a particle to an absorbing boundary by two different methods: first, using a Fokker-Planck description of the system and then using a general approach to first-passage under restart. Closed-form formul\ae \;were obtained for the Laplace transform of the propagator governing this process and for the distribution of the FPT to the absorbing boundary. Exact expressions for the mean FPT and the restart rate that minimizes it were also obtained. For positive drift velocities, this optimal restart rate was found to transition from a value of zero attained at the drift-controlled regime to a value of $r^{\star}\simeq r_0^{\star}\left(1-Pe\right)$ attained at the diffusion-controlled regime. The speedup conferred by optimal restart also undergoes a transition. In the diffusion-controlled regime optimal restart can reduce the mean FPT by a factor of $\sim1/Pe$, which can be huge for small $Pe$. However, in the drift-controlled regime the optimal restart rate is zero, i.e., restart cannot accelerate the process at all.  Another hallmark of the transition is observed when examining relative fluctuations of the FPT under optimal restart. These deviate from a universal value of unity as one transitions from the diffusion-controlled regime to the drift-controlled regime. 

Taken together, our results provide a comprehensive quantification of the effect stochastic restart has on drift-diffusion. It also suggests drift-diffusion as a concrete model system where first-passage under restart can be experimentally realized to demonstrate the restart transition. Drift-diffusion can be realized experimentally, e.g., with a colloidal particle in a flow chamber. Such setup was recently used as part of an experimental demonstration of an information machine \cite{Yael}. There, a barrier made of light was used to prevent the particle from being carried away by the flow, but light can also be used to trap and manipulate the particle. Indeed, optical tweezers can be used to return the particle to its initial position whenever it fails to hit the boundary within a given time window that could, in principle, be random. Experimental realization of this setup could thus be used to demonstrate some of the interesting physics that was predicted above, and in particular serve to show that in the diffusion-controlled regime optimal restart sets the relative standard deviation of the FPT to unity \cite{ReuveniPRL}. This would be the first experimental verification of this universal principle. 

\begin{figure}[t!]
\begin{centering}
\includegraphics[width=7.0cm]{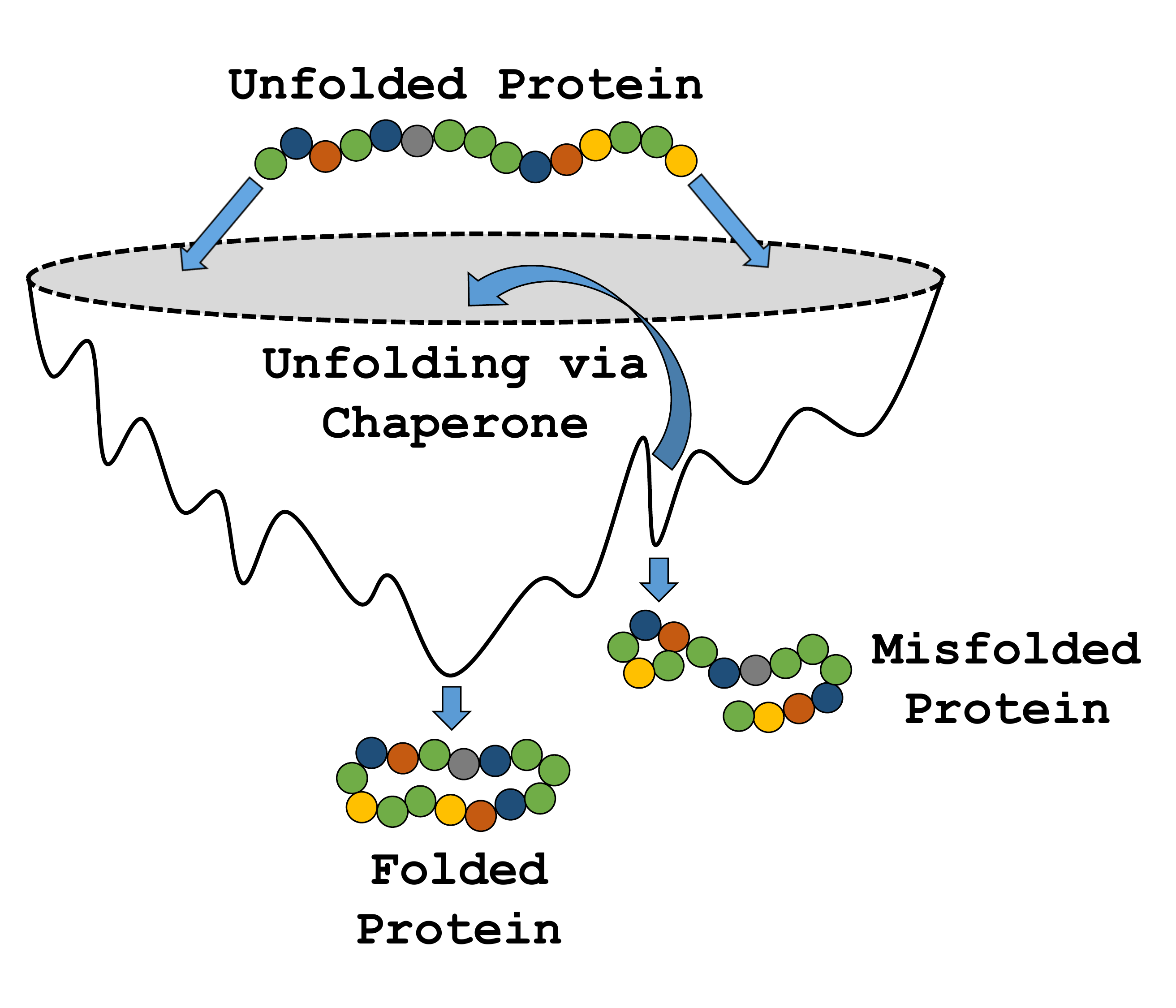}
\end{centering}
\caption{A schematic of protein folding envisioned as diffusion in a rugged potential landscape. Chaperones save misfolded proteins by restarting the folding process, but the connection with the problem of first-passage under restart has so far been overlooked. }
\label{Fig8}
\end{figure}

Finally, we note that the approach taken herein can also be applied to diffusion in higher dimensions and in complicated potential landscapes. Strong motivation to do so comes e.g., from the problem of protein folding (Fig.~\ref{Fig8}). There, the set of unfolded states are high in energy, and the natively folded state is depicted as the global minimum of the energy function. Meta-stable states, in which the protein is only partly folded, also exist and are represented by local minima that are scattered across the energy landscape. A problem then arises: some of these local minima are deep enough to form long-lived meta-stable states which pose a grave danger to living cells as partially-folded proteins are not only dysfunctional but also tend to aggregate into harmful clusters. Enter chaperones---a group of molecular machines whose job in the cell is to prevent, or reverse, protein misfolding. In particular, by using energy some chaperones act as unfolders that restart the folding process, thus giving misfolds another chance to find the native  state. 

Situations of the type described above clarify why one would want to study how restart affects the FPT---to a marked target---of a particle (reaction coordinate) diffusing in a rugged potential landscape. The general theory of first-passage under restart then asserts that the presence of a restarting agent, e.g., a molecular chaperone, would be favorable whenever stochastic fluctuations in the FPT are large by the $CV$ measure \cite{Restart-Biophysics1,Restart-Biophysics2,ReuveniPRL,PalReuveniPRL}. For example, in protein folding large fluctuations in the FPT to the native state can arise when e.g., most folding attempts successfully come to completion after some typical time $\tau_1$ but a small fraction requires a time $\tau_2\gg\tau_1$ as a result of prolonged trapping in meta-stable states. Restarting ongoing folding processes at a rate $\tau_2^{-1}\ll r \ll\tau_1^{-1}$ is then expected to be highly beneficial; but the effect of restart could be easily inverted as changes in pH or temperature alter the ratio between $\tau_1$ and $\tau_2$ and thus the $CV$ of the folding process. To study this, and many other similar problems, one can apply the same methodology that was utilized in Sec. II.B. First, define the first-passage time problem in the absence of resetting and solve it (or get the solution from the literature in case it is known). Then, substitute this solution into Eqs. (\ref{eq:fptd_r}) and (\ref{eq:mfpt_lt}) to get the solution to the problem in the presence of resetting. As parameters are varied in the system, a transition in the effect of restart will occur when the $CV$ of the restart-free first-passage time crosses the value of unity (this transition point is universal). Here, we analysed Eq. (\ref{eq:mfpt_lt})  and showed that the P\'eclet number governs this transition for drift-diffusion. Equation  (\ref{eq:mfpt_lt}) can also be analyzed for more complicated situations, but the challenge is then to identify a P\'eclet like quantity that governs the restart transition near and far from the critical point. This challenge will be addressed elsewhere.
\section{Acknowledgments}
S. Ray acknowledges support from the Raymond and Beverly Sackler Center for Computational Molecular and Materials Science, Tel Aviv University. S. Reuveni acknowledges support from the Azrieli Foundation. D. Mondal acknowledges the Ratner Center for Single Molecule Science for supporting his visit to the School of Chemistry at Tel Aviv University. All authors thank Yair Shokef, Arnab Pal, Yuval Scher, Sarah Kostinski and Ofek Lauber for reading and commenting on early versions of this manuscript.
\appendix
\numberwithin{equation}{section}
\setcounter{equation}{0}
\section{Appendix A: Derivation of Eq.~(\ref{eq:pxs})}
\renewcommand{\theequation}{A.\arabic{equation}}
Eq.~(\ref{eq:lt}) in the main text is a second-order, linear, non-homogeneous differential equation. It has general spatial coordinate-dependent solution
\begin{eqnarray}
\begin{aligned}
\tilde{p}_{+}(x,s)&=A_1(s)e^{\alpha_+ x}+B_1(s)e^{\alpha_- x}\;\;\;\text{if}\;\;\;0\leq{x}\leq{L} \\
\tilde{p}_{-}(x,s)&=A_2(s)e^{\alpha_+ x}+B_2(s)e^{\alpha_- x}\;\;\text{if}\;\;-\infty<x\leq{0}.
\end{aligned}
\label{eq:A1}
\end{eqnarray}
Here $\alpha_{\pm} =\frac{1}{2D}\left[V\pm\sqrt{V^2+4D(s+r)}\right]$ are the two roots of the characteristic equation associated with the homogeneous differential equation
\begin{align}
\dfrac{\partial^2 \tilde{p}(x,s)}{\partial x^2}-\left(\frac{V}{D}\right)\dfrac{\partial \tilde{p}(x,s)}{\partial x}-\left(\frac{s+r}{D}\right)\tilde{p}(x,s)=0.
\label{eq:A2}
\end{align}
Since $V$ is real and $D$, $s$ and $r$ all are real and positive, it is evident from the expressions of $\alpha_{\pm}$ that $\alpha_+>0$ while $\alpha_-<0$. In order to find a specific solution to Eq.~(\ref{eq:lt}), we need to calculate $A_1(s)$, $A_2(s)$, $B_1(s)$ and $B_2(s)$ explicitly. We can accomplish that in the following way.\\
\indent
To prevent $\tilde{p}_{-}(x,s)$ from diverging at $x\rightarrow-\infty$, we set $B_2(s)=0$. The other boundary condition, $\tilde{p}_{+}(L,s)=0$, gives 
\begin{equation}
A_1(s)e^{\alpha_+L}+B_1(s)e^{\alpha_-L}=0.
\label{eq:A3}
\end{equation}
In addition, $\tilde{p}_{-}(x,s)$ should be continuous at $x=0$, i.e., $\tilde{p}_{+}(0,s)=\tilde{p}_{-}(0,s)$, leading to 
\begin{equation}
A_1(s)+B_1(s)=A_2(s).
\label{eq:A4}
\end{equation}
The fourth and final condition can be obtained by integrating Eq.~(\ref{eq:lt}) in the main text over the narrow spatial interval [$-\epsilon,\epsilon$]. Doing so, we get
\begin{eqnarray}
\left[\dfrac{\partial \tilde{p}(x,s)}{\partial x}\right]_{-\epsilon}^{+\epsilon}&-&\left(\frac{V}{D}\right)\left[\tilde{p}(x,s)\right]_{-\epsilon}^{+\epsilon}-\left(\frac{s+r}{D}\right)\int_{-\epsilon}^{+\epsilon}\tilde{p}(x,s)dx \nonumber \\
&=&\frac{-1-r \tilde{Q}(s)}{D},
\label{eq:A5}
\end{eqnarray}
where we have utilized the identity $\int_{-\epsilon}^{+\epsilon}\delta(x)dx=1$. Due to the continuity of $\tilde{p}(x,s)$ at $x=0$, $\lim_{\epsilon\rightarrow0}\left[\tilde{p}(x,s)\right]_{-\epsilon}^{+\epsilon}=0$. In addition, $\lim_{\epsilon\rightarrow0}\int_{-\epsilon}^{+\epsilon}\tilde{p}(x,s)dx=0$. Thus in the limit of $\epsilon\rightarrow0$, Eq.~(\ref{eq:A5}) gives
\begin{equation}
\frac{\partial}{\partial x}\left[\tilde{p}_{+}(x,s)-\tilde{p}_{-}(x,s)\right]_{x=0}=\frac{-1-r \tilde{Q}(s)}{D},
\label{eq:A6}
\end{equation}
i.e., a finite discontinuity in the first derivative. Plugging in Eq.~(\ref{eq:A1}) into Eq.~(\ref{eq:A6}) we obtain
\begin{equation}
\alpha_+A_1(s)+\alpha_-B_1(s)=\alpha_+A_2(s)-\left[\frac{1+r \tilde{Q}(s)}{D}\right].
\label{eq:A7}
\end{equation}
Solving Eqs.(\ref{eq:A3}), (\ref{eq:A4}), and (\ref{eq:A7}), we get
\begin{eqnarray}
\begin{aligned}
A_1(s)&=e^{(\alpha_--\alpha_+) L}\left[\frac{1+r \tilde{Q}(s)}{D(\alpha_--\alpha_+)}\right]\\
A_2(s)&=\left(e^{(\alpha_--\alpha_+) L}-1\right)\left[\frac{1+r \tilde{Q}(s)}{D(\alpha_--\alpha_+)}\right]\\
B_1(s)&=-\left[\frac{1+r \tilde{Q}(s)}{D(\alpha_--\alpha_+)}\right].
\end{aligned}
\label{eq:A8}
\end{eqnarray}
Plugging in Eq.~(\ref{eq:A8}) into Eq.~(\ref{eq:A1}), we obtain Eq.~(\ref{eq:pxs}) in the main text.
\section{Appendix B: Optimal restart rate for $V<0$}
\renewcommand{\theequation}{B.\arabic{equation}}
\begin{figure}[t!]
\begin{centering}
\includegraphics[width=7.8cm]{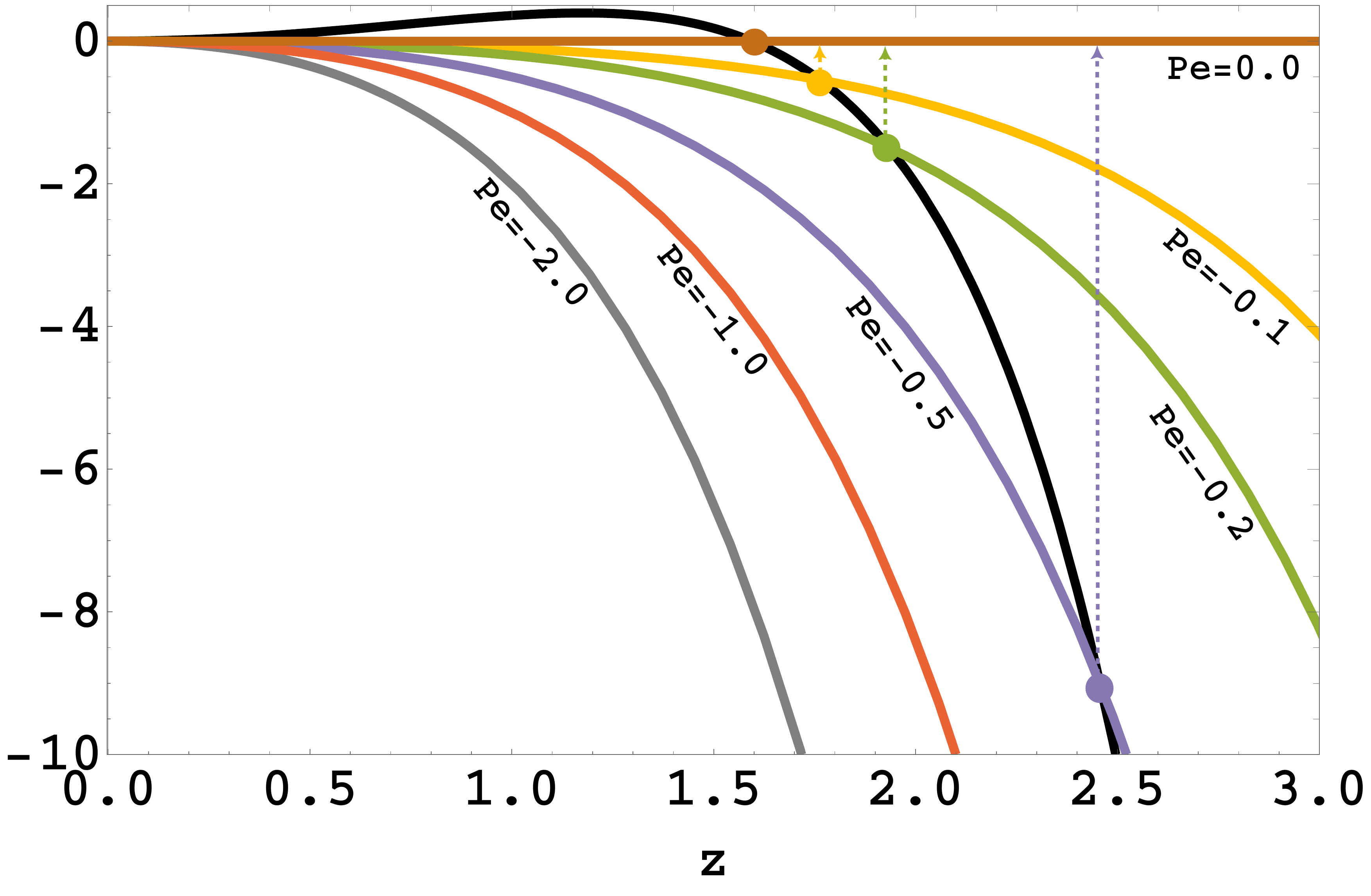}
\end{centering}
\caption{Left hand side (colored lines) and right hand side (black line) of Eq.~(\ref{eq:trans}) vs. the reduced variable $z$ from \eref{eq:z_def} for negative P\'eclet numbers. Colored circles denote solutions to \eref{eq:trans} for different $Pe$.}
\label{Fig9}
\end{figure}
Analysis in the main text was focused on positive drift velocities ($V>0$). However, results for $V<0$ can be obtained by following a similar  procedure since \eref{eq:fme} to \eref{eq:optr_Pe} hold for negative drift velocities as well. When $V<0$, the drive is directed away from the absorbing boundary and the particle has a non-zero probability to escape to infinity, which means that the mean FPT necessarily diverges. Introducing restart renders the mean completion time finite [\eref{eq:mfpt}], thereby leading to infinite speedup. Therefore, the effect of restart in the $V<0$ regime is one and the same: it always expedites the process. No restart transition is thus expected when the drift velocity is negative.
\begin{figure}[t!]
\begin{centering}
\includegraphics[width=7.8cm]{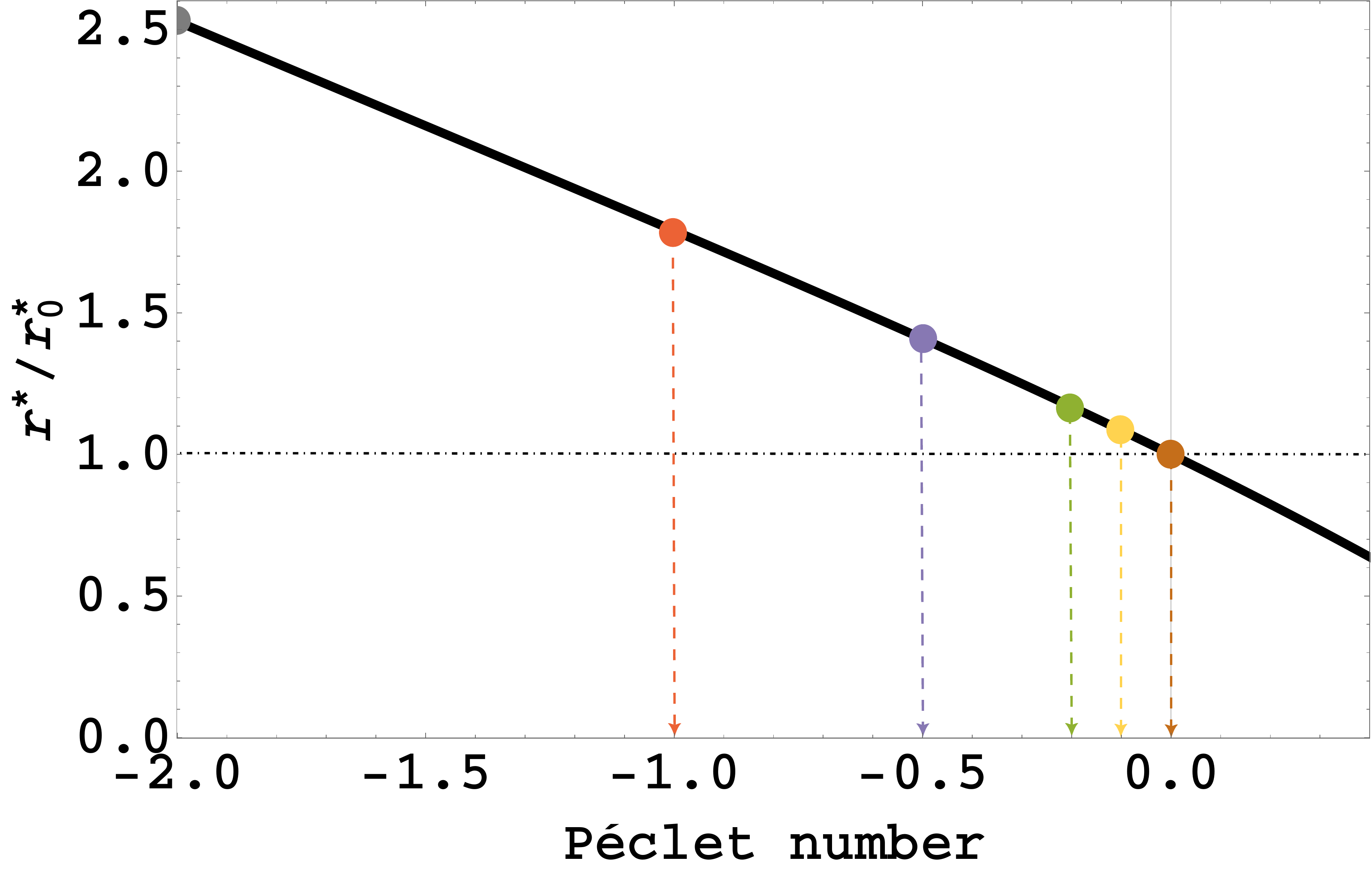}
\end{centering}
\caption{The scaled optimal restart rate from \eref{eq:optr_Pe} vs. the P\'eclet number from \eref{eq:Pe} (solid black line). Colored circles correspond to $Pe$ values from Fig.~\ref{Fig9}. }
\label{Fig10}
\end{figure}

For negative drift velocities the P\'eclet number is negative, i.e., $Pe\in(-\infty,0)$. In order to calculate the optimal restart rate in this case, we first solve \eref{eq:trans} with $Pe<0$ in a manner similar to that described in the main text. In Fig. \ref{Fig9}, we plot together the left-hand side (colored lines denoting different values of the P\'eclet number) and right-hand side (black line) of \eref{eq:trans}. Solutions, denoted by $z^{\star}$, correspond to $z$ values at which the two curves intersect. The optimal restart rates are calculated from these $z^{\star}$ values following \eref{eq:optr_Pe}. We plot the scaled optimal restart rates in Fig. \ref{Fig10} to observe that it monotonically increases with the magnitude of the negative P\'eclet number. 

\end{document}